\documentclass[10pt]{article}
\usepackage{graphicx}
\usepackage{amsmath}
\usepackage{amssymb}
\usepackage{caption2}
\begin{document}
\begin{center}
{\large {\bf \sc{   Analysis of the triply-heavy baryon states   with the  QCD sum rules
  }}} \\[2mm]
Zhi-Gang  Wang \footnote{E-mail: zgwang@aliyun.com.  }   \\
 Department of Physics, North China Electric Power University, Baoding 071003, P. R. China
\end{center}

\begin{abstract}
In this article, we restudy the mass spectrum of the ground state triply-heavy baryon states with the QCD sum rules by carrying  out the operator product expansion up to the vacuum condensates of dimension 6 in a consistent way and preforming a novel  analysis. It is the first time to take into account the three-gluon condensates in the QCD sum rules for the triply-heavy baryon states.
\end{abstract}



\section{Introduction}
In  recent  years,   a large number of  heavy baryon states, charmonium-like states and bottomonium-like states  have been observed,  and have attracted   intensive attentions and have revitalized many works   on the singly-heavy, doubly-heavy, triply-heavy and quadruply-heavy hadron  spectroscopy \cite{PDG}.
In 2017, the LHCb collaboration observed the doubly-charmed baryon state  $\Xi_{cc}^{++}$ in the $\Lambda_c^+ K^- \pi^+\pi^+$ mass spectrum for the first time \cite{LHCb-Xicc}, while
the doubly-charmed baryon states $\Xi_{cc}^{+}$ and $\Omega_{cc}^{+}$ are still unobserved.
In 2020, the LHCb  collaboration studied the $J/\psi J/\psi$   invariant mass distribution using $pp$ collision data at center-of-mass energies of $\sqrt{s}=$7, 8  and 13 TeV, and observed a narrow resonance structure $X(6900)$ around $6.9\, \rm{GeV}$  and   a broad structure just above the $J/\psi J/\psi$ mass with global significances of more than $5\sigma$ \cite{LHCb-cccc-2006}. They are good candidates for the fully-charmed tetraquark states, also are the first fully-heavy exotic multiquark candidates claimed  experimentally to date. If they are really fully-charmed tetraquark states, we have observed doubly-charmed and quadruply-charmed hadrons, but  no experimental evidences  for the triply-charmed hadrons.  The observation of the $\Xi_{cc}^{++}$ and $X(6900)$ provides some crucial experimental inputs  on the strong correlation between the two charm quarks, which maybe  shed light on the
  spectroscopy of the doubly-heavy, triply-heavy baryon states, doubly-heavy, triply-heavy, quadruply-heavy tetraquark states and pentaquark states.
  On the other hand, the spectrum of the triply-heavy baryon states have been studied extensively via different theoretical approaches, such  as the lattice QCD \cite{LQCD-1,LQCD-2,LQCD-3,LQCD-4,LQCD-5,LQCD-6,LQCD-7,LQCD-8}, the QCD sum rules \cite{QCDSR-1,QCDSR-2,QCDSR-3,QCDSR-4}, various potential quark models \cite{PQM-1,PQM-2,PQM-3,PQM-4,PQM-5,PQM-6,PQM-7,PQM-8,PQM-9,PQM-10,PQM-11,PQM-12,PQM-13,PQM-14,PQM-15,PQM-16}, Fadeev equation \cite{Faddeev-1,Faddeev-2,Faddeev-3,Faddeev-4}, the Regge trajectories \cite{Regge-1,Regge-2}, etc.  The predicted triply-heavy baryon masses vary in a rather large range, more theoretical works are still needed for the sake of obtaining more precise inputs  for comparing to the  experimental data in the future.

 The QCD sum rules approach is a  powerful theoretical tool in studying the mass spectrum of the heavy flavor hadrons, and  plays  an important role in assigning the new baryon states, exotic tetraquark (molecular) states and pentaquark (molecular) states.  The ground state triply-heavy baryon states $QQQ$ and $QQQ^\prime$ have been studied with the QCD sum rules by taking into account the perturbative contributions and the gluon condensate contributions in performing the operator product expansion \cite{QCDSR-1,QCDSR-2,QCDSR-3,QCDSR-4}. In calculations, the constant or universal heavy-quark pole masses or $\overline{MS}$ masses are chosen in all the channels \cite{QCDSR-1,QCDSR-2,QCDSR-3,QCDSR-4}. In this article, we restudy
  the mass spectrum of the ground state  triply-heavy baryon states by taking into account the three-gluon condensates, it is the first time to take into account  the three-gluon condensates in the QCD sum rules for the triply-heavy baryon states. Furthermore, we pay special attentions to the heavy quark masses, and choose the values which work well in studying the doubly-heavy baryon states \cite{Wang-cc-baryon-penta}, hidden-charm tetraquark states \cite{WangZG-Hiddencharm, WangZG-Vector-tetra}, hidden-bottom tetraquark states \cite{ WangZG-Hiddenbottom}, hidden-charm pentaquark states \cite{WangZG-hiddencharm-penta}, fully-charmed tetraquark states \cite{WangZG-QQQQ}, and perform a novel analysis.

The article is arranged as follows:  we obtain  the QCD sum rules for the masses and pole residues of  the
 triply-heavy baryon   states in Sect.2;  in Sect.3, we present the numerical results and discussions; and Sect.4 is reserved for our
conclusion.

\section{QCD sum rules for  the triply-heavy baryon  states}

Firstly, we write down  the two-point correlation functions  $\Pi(p)$ and $\Pi_{\mu\nu}(p)$   in the QCD sum rules,
\begin{eqnarray}
\Pi(p)&=&i\int d^4x e^{ip \cdot x} \langle0|T\left\{J(x)\bar{J}(0)\right\}|0\rangle \, , \nonumber \\
\Pi_{\mu\nu}(p)&=&i\int d^4x e^{ip \cdot x} \langle0|T\left\{J_\mu(x)\bar{J}_{\nu}(0)\right\}|0\rangle \, ,
\end{eqnarray}
where $J(x)=J^{QQQ'}(x)$, $J_\mu(x)=J^{QQQ'}_\mu(x)$, $J^{QQQ}_\mu(x)$,
\begin{eqnarray}
J^{QQQ'}(x)&=& \varepsilon^{ijk}  Q^T_i(x)C\gamma_\mu Q_j(x) \gamma^\mu \gamma_5   Q'_k(x) \, , \nonumber \\
J^{QQQ'}_\mu(x)&=& \varepsilon^{ijk}  Q^T_i(x)C\gamma_\mu Q_j(x)    Q'_k(x) \, , \nonumber \\
J^{QQQ}_\mu(x)&=& \varepsilon^{ijk}  Q^T_i(x)C\gamma_\mu Q_j(x)    Q_k(x) \, ,
\end{eqnarray}
where   $Q$, $Q'=b$, $c$, $Q\neq Q^\prime$, the $i$, $i$ and $k$ are color indexes, and the $C$ is the charge conjugation
matrix.
We choose the Ioffe-type currents $J(x)$ and $J_\mu(x)$ to interpolate the triply-heavy baryon states with the spin-parity $J^P={\frac{1}{2}}^+$ and ${\frac{3}{2}}^+$, respectively,
\begin{eqnarray}
\langle 0|J(0)|\Omega_{QQQ^\prime,+}(p)\rangle&=&\lambda_{+} U_{+}(p,s)\, , \nonumber\\
\langle 0|J_\mu(0)|\Omega_{QQQ^{(\prime)},+}^*(p)\rangle&=&\lambda_{+} U^{+}_\mu(p,s)\, ,
\end{eqnarray}
where the $\Omega_{QQQ^\prime,+}$ and $\Omega^*_{QQQ^{(\prime)},+}$ represent the triply-heavy baryon states with the spin-parity $J^P={\frac{1}{2}}^+$ and ${\frac{3}{2}}^+$, respectively, the $\lambda_{+}$  are the pole residues, the $U(p,s)$ and $U_\mu(p,s)$ are the Dirac spinors, the subscript or superscript  $+$ denotes the parity.
 The currents $J(x)$ and $J_\mu(x)$ also couple potentially to the negative-parity triply-heavy baryon states $\Omega_{QQQ^\prime,-}$ and $\Omega^*_{QQQ^{(\prime)},-}$, respectively,
because multiplying $i \gamma_{5}$ to the currents $J(x)$  and $J_\mu(x)$ changes their
parity \cite{QCDSR-4,Wang-cc-baryon-penta,WangZG-hiddencharm-penta,Oka96,WangHbaryon},
\begin{eqnarray}
\langle 0|J(0)|\Omega_{QQQ^\prime,-}(p)\rangle&=&\lambda_{-} i\gamma_5 U_{-}(p,s)\, , \nonumber\\
\langle 0|J_\mu(0)|\Omega_{QQQ^{(\prime)},-}^*(p)\rangle&=&\lambda_{-} i\gamma_5 U^{-}_\mu(p,s)\, ,
\end{eqnarray}
again the subscript or superscript  $-$ denotes the parity. On the other hand, we can use the valance quarks and spin-parity to represent the triply-heavy baryon states, for example, $QQQ^\prime({\frac{3}{2}}^+)$. We cannot construct  the currents $J^{QQQ}(x)= \varepsilon^{ijk}  Q^T_i(x)C\gamma_\mu Q_j(x) \gamma^\mu \gamma_5   Q_k(x) $ to interpolate the  triply-heavy baryon states $\Omega_{QQQ,+}$ with the spin-parity $J^{P}={\frac{1}{2}}^+$, because such current operators  cannot exist due to the Fermi-Dirac statistics.

We  insert  a complete set  of intermediate triply-heavy baryon states with the same quantum numbers as the current operators $J(x)$, $i\gamma_5 J(x)$,
$J_\mu(x)$ and $i\gamma_5J_\mu(x)$ into the correlation functions $\Pi(p)$ and
$\Pi_{\mu\nu}(p)$ to obtain the hadron representation
\cite{SVZ79,Reinders85}. After isolating the pole terms of the lowest
states of the positive-parity and negative-parity  triply-heavy baryon states, we obtain the
 results:
\begin{eqnarray}
 \Pi(p) & = &\lambda_+^2 {\!\not\!{p} + M_{+} \over M^{2}_+ -p^{2} } + \lambda_{-}^2 {\!\not\!{p} - M_{-} \over M_{-}^{2}-p^{2} } + \cdots \, , \nonumber \\
   &=&  \Pi_1(p^2) \!\not\!{p} +\Pi_0(p^2) \, ,\nonumber \\
 \Pi_{\mu\nu}(p)&=&\Pi(p)\left( -g_{\mu\nu}+\cdots \right)+\cdots \, ,
    \end{eqnarray}
  we choose the tensor structure $g_{\mu\nu}$ to study the spin $J=\frac{3}{2}$ triply-heavy baryon states.

 We can obtain the hadronic spectral densities $\rho^1_H(s)$  and $\rho^0_H(s)$  at the hadron side through dispersion relation,
\begin{eqnarray}
\rho^1_H(s)&=& \frac{{\rm Im} \, \Pi_1(s)}{\pi} \nonumber\\
 & = & \lambda_+^2 \, \delta\left(s - M_{+}^2\right)+    \lambda_{-}^{2}  \, \delta\left(s - M_{-}^2\right) \, , \nonumber \\
\rho^0_H(s)&=&  \frac{{\rm Im} \, \Pi_0(s)}{\pi}\nonumber\\
 & = & M_{+}\lambda_+^2 \, \delta\left(s - M_{+}^2\right)-M_{-}    \lambda_{-}^{2}  \, \delta\left(s - M_{-}^2\right) \, ,
\end{eqnarray}
then introduce the weight function $\exp\left(-\frac{s}{T^2} \right)$, and obtain the QCD sum rules at the hadron side,
\begin{eqnarray}
 2M_{+}\lambda_{+}^2 \exp\left( -\frac{M_{+}^2}{T^2}\right) & = & \int_{\Delta^2}^{s_0}ds \left[\sqrt{s}\rho^1_H(s)+ \rho^0_H(s)\right] \exp\left(-\frac{s}{T^2} \right)\ \, ,
\end{eqnarray}
where the thresholds $\Delta^2=(2m_Q+m_{Q'})^2$ or  $9m^2_Q$, the $T^2$ are the Borel parameters, and the $s_0$ are the continuum threshold parameters.
The combinations   $\sqrt{s}\rho^1_H(s)+ \rho^0_H(s)$ and $\sqrt{s}\rho^1_H(s)- \rho^0_H(s)$ contain the
contributions  from the positive-parity   and negative-parity triply-heavy baryon states,  respectively.

 Now  we  briefly outline the operator product expansion performed  at the deep Euclidean  region $p^2 \ll 0$.
 We contract the heavy quark fields in the correlation functions $\Pi(p)$ and $\Pi_{\mu\nu}(p)$ with
Wick theorem,  substitute the full  heavy quark propagators $S_{ij}(x)$ into the
correlation functions  $\Pi(p)$ and $\Pi_{\mu\nu}(p)$ firstly,
\begin{eqnarray}
S_{ij}(x)&=&\frac{i}{(2\pi)^4}\int d^4k e^{-ik \cdot x} \left\{
\frac{\delta_{ij}}{\!\not\!{k}-m_Q}
-\frac{g_sG^n_{\alpha\beta}t^n_{ij}}{4}\frac{\sigma^{\alpha\beta}(\!\not\!{k}+m_Q)+(\!\not\!{k}+m_Q)
\sigma^{\alpha\beta}}{(k^2-m_Q^2)^2}\right.\nonumber\\
&& -\frac{g_s^2 (t^at^b)_{ij} G^a_{\alpha\beta}G^b_{\mu\nu}(f^{\alpha\beta\mu\nu}+f^{\alpha\mu\beta\nu}+f^{\alpha\mu\nu\beta}) }{4(k^2-m_Q^2)^5}\nonumber\\
&&\left.+\frac{\langle g_s^3GGG\rangle}{48}\frac{(\!\not\!{k}+m_Q)\left[\!\not\!{k}(k^2-3m_Q^2)+2m_Q(2k^2-m_Q^2) \right](\!\not\!{k}+m_Q)}{(k^2-m_Q^2)^6}+\cdots\right\}\, ,\nonumber\\
f^{\alpha\beta\mu\nu}&=&(\!\not\!{k}+m_Q)\gamma^\alpha(\!\not\!{k}+m_Q)\gamma^\beta(\!\not\!{k}+m_Q)\gamma^\mu(\!\not\!{k}+m_Q)\gamma^\nu(\!\not\!{k}+m_Q)\, ,
\end{eqnarray}
 where $\langle g_s^3GGG\rangle=\langle g_s^3f_{abc}G^a_{\mu\nu}G_b^{\nu\alpha}G^b_{\alpha}{}^\mu\rangle$, $t^n=\frac{\lambda^n}{2}$, the $\lambda^n$ is the Gell-Mann matrix, the $i$, $j$ are the color indexes \cite{Reinders85},
then  complete  the integrals in the coordinate space and momentum
space sequentially to obtain the correlation functions $\Pi(p)$ and $\Pi_{\mu\nu}(p)$ at the quark-gluon level, finally we obtain
the corresponding QCD spectral densities through dispersion relation,
\begin{eqnarray}
\rho^1_{QCD}(s)&=& \frac{{\rm Im} \, \Pi_1(s)}{\pi}  \, , \nonumber \\
\rho^0_{QCD}(s)&=&  \frac{{\rm Im} \, \Pi_0(s)}{\pi} \, .
\end{eqnarray}

We  match the hadron side with the QCD side of the correlation functions $\Pi_1(p^2)$ and $\Pi_0(p^2)$ below the continuum thresholds $s_0$,  introduce the weight function $\exp\left(-\frac{s}{T^2} \right)$, and obtain the QCD sum rules,
\begin{eqnarray}\label{QCDSR}
 2M_{+}\lambda_{+}^2 \exp\left( -\frac{M_{+}^2}{T^2}\right) & = & \int_{\Delta^2}^{s_0}ds \left[\sqrt{s}\rho^1_H(s)+ \rho^0_H(s)\right] \exp\left(-\frac{s}{T^2} \right)\ \, , \nonumber\\
 & = & \int_{\Delta^2}^{s_0}ds \left[\sqrt{s}\rho^1_{QCD}(s)+ \rho^0_{QCD}(s)\right] \exp\left(-\frac{s}{T^2} \right)\ \, ,
\end{eqnarray}
where $\rho^1_{QCD}(s)=\rho^1_{QQQ,\frac{3}{2}}(s)$, $\rho^1_{QQQ^\prime,\frac{3}{2}}(s)$, $\rho^1_{QQQ^\prime,\frac{1}{2}}(s)$,
$\rho^0_{QCD}(s)=m_Q\rho^0_{QQQ,\frac{3}{2}}(s)$, $m_{Q^\prime}\rho^0_{QQQ^\prime,\frac{3}{2}}(s)$, $m_{Q^\prime}\rho^0_{QQQ^\prime,\frac{1}{2}}(s)$,
\begin{eqnarray}
\rho^1_{QQQ,\frac{3}{2}}(s)&=&\frac{3}{64\pi^4}\int_{z_i}^{z_f}dz \int_{y_i}^{y_f}dy \, yz(1-y-z) (s-\widetilde{m}_{Q}^2)(11s-5\widetilde{m}_{Q}^2) \nonumber\\
&&+\frac{15m_Q^2}{32\pi^4}\int_{z_i}^{z_f}dz \int_{y_i}^{y_f}dy \,y\, (s-\widetilde{m}_Q^2)\nonumber\\
&&-\frac{m_Q^2}{32\pi^2}\langle\frac{\alpha_{s}GG}{\pi}\rangle\int_{z_i}^{z_f}dz \int_{y_i}^{y_f}dy \, \frac{z(1-y-z)}{y^{2}}  \left(1+\frac{3s}{2T^2}\right)\delta\left(s-\widetilde{m}_{Q}^2\right)\nonumber\\
&&-\frac{5m_Q^4}{192\pi^2T^2}\langle\frac{\alpha_{s}GG}{\pi}\rangle\int_{z_i}^{z_f}dz \int_{y_i}^{y_f}dy \, \frac{1}{y^{2}} \delta\left(s-\widetilde{m}_{Q}^2\right)\nonumber\\
&&-\frac{5m_Q^4}{96\pi^2T^2}\langle\frac{\alpha_{s}GG}{\pi}\rangle\int_{z_i}^{z_f}dz \int_{y_i}^{y_f}dy \, \frac{z}{y^{3}} \delta\left(s-\widetilde{m}_{Q}^2\right)\nonumber\\
&&+\frac{5m_Q^2}{32\pi^2}\langle\frac{\alpha_{s}GG}{\pi}\rangle\int_{z_i}^{z_f}dz \int_{y_i}^{y_f}dy \, \frac{z}{y^{2}} \delta\left(s-\widetilde{m}_{Q}^2\right)\nonumber\\
&&-\frac{25}{384\pi^2}\langle\frac{\alpha_{s}GG}{\pi}\rangle\int_{z_i}^{z_f}dz \int_{y_i}^{y_f}dy \, (1-y-z)\left[ 1+\frac{7s}{25}\delta\left(s-\widetilde{m}_{Q}^2\right)\right]\nonumber\\
&&-\frac{5m_Q^2}{192\pi^2}\langle\frac{\alpha_{s}GG}{\pi}\rangle\int_{z_i}^{z_f}dz \int_{y_i}^{y_f}dy \, \frac{1}{z} \delta\left(s-\widetilde{m}_{Q}^2\right)\nonumber\\
&&-\frac{m_Q^2\langle g_s^3GGG\rangle}{512\pi^4T^2}\int_{z_i}^{z_f}dz \int_{y_i}^{y_f}dy \, \frac{z(1-y-z)}{y^{3}}\left( 1-\frac{3s}{T^2}\right)\delta\left(s-\widetilde{m}_{Q}^2\right)\nonumber\\
&&+ \frac{5m_Q^4\langle g_s^3GGG\rangle}{768\pi^4T^4}\int_{z_i}^{z_f}dz \int_{y_i}^{y_f}dy \, \frac{z}{y^{4}}\delta\left(s-\widetilde{m}_{Q}^2\right)\nonumber\\
&&+\frac{5m_Q^4\langle g_s^3GGG\rangle}{1536\pi^4T^4} \int_{z_i}^{z_f}dz \int_{y_i}^{y_f}dy \, \frac{1}{y^{3}}\delta\left(s-\widetilde{m}_{Q}^2\right)\nonumber\\
&&- \frac{\langle g_s^3GGG\rangle}{1024\pi^4}\int_{z_i}^{z_f}dz \int_{y_i}^{y_f}dy \, \frac{z(1-y-z)}{y^{2}}\left( 2+\frac{3s}{T^2}\right)\delta\left(s-\widetilde{m}_{Q}^2\right)\nonumber\\
&&- \frac{5m_Q^2\langle g_s^3GGG\rangle}{256\pi^4T^2} \int_{z_i}^{z_f}dz \int_{y_i}^{y_f}dy \, \frac{z}{y^{3}}\delta\left(s-\widetilde{m}_{Q}^2\right)\nonumber\\
&&- \frac{23m_Q^2\langle g_s^3GGG\rangle}{4608\pi^4T^2} \int_{z_i}^{z_f}dz \int_{y_i}^{y_f}dy \, \frac{1}{y^{2}}\delta\left(s-\widetilde{m}_{Q}^2\right)\nonumber\\
&&+\frac{11m_Q^2\langle g_s^3GGG\rangle}{4608\pi^4T^2}\int_{z_i}^{z_f}dz \int_{y_i}^{y_f}dy \, \frac{1-y-z}{y^{2}}\left( 1+\frac{7s}{11T^2}\right)\delta\left(s-\widetilde{m}_{Q}^2\right)\nonumber\\
&&+ \frac{m_Q^2\langle g_s^3GGG\rangle}{2304\pi^4T^2}\int_{z_i}^{z_f}dz \int_{y_i}^{y_f}dy \, \frac{1}{y}\left(1+\frac{s}{T^2}\right)\delta\left(s-\widetilde{m}_{Q}^2\right)\nonumber\\
&&+ \frac{m_Q^2\langle g_s^3GGG\rangle}{768\pi^4T^2} \int_{z_i}^{z_f}dz \int_{y_i}^{y_f}dy \, \frac{1-y-z}{zy^{2}}\delta\left(s-\widetilde{m}_{Q}^2\right)\nonumber\\
&&- \frac{m_Q^2\langle g_s^3GGG\rangle}{384\pi^4T^2}  \int_{z_i}^{z_f}dz \int_{y_i}^{y_f}dy \, \frac{1-y-z}{zy}
\left( 1-\frac{s}{3T^2}\right)\delta\left(s-\widetilde{m}_{Q}^2\right)\nonumber\\
&&+ \frac{m_Q^4\langle g_s^3GGG \rangle}{1536\pi^4T^4} \int_{z_i}^{z_f}dz \int_{y_i}^{y_f}dy \, \frac{1}{y^{3}}\delta\left(s-\widetilde{m}_{Q}^2\right) \nonumber
\end{eqnarray}

\begin{eqnarray}
&&+ \frac{5m_Q^4\langle g_s^3 GGG\rangle}{4608\pi^4T^4}  \int_{z_i}^{z_f}dz \int_{y_i}^{y_f}dy \, \frac{1}{zy^{2}}\delta\left(s-\widetilde{m}_{Q}^2\right)\nonumber\\
&&+ \frac{m_Q^4\langle g_s^3 GGG\rangle}{1536\pi^4T^4} \int_{z_i}^{z_f}dz \int_{y_i}^{y_f}dy \, \frac{1-y-z}{zy^{3}}\delta\left(s-\widetilde{m}_{Q}^2\right)\nonumber\\
&&- \frac{\langle g_s^3 GGG \rangle}{512\pi^4} \int_{z_i}^{z_f}dz \int_{y_i}^{y_f}dy \, \frac{1-y-z}{y}\left( 3+\frac{7s}{6T^2}\right)\delta\left(s-\widetilde{m}_{Q}^2\right)\nonumber\\
&&- \frac{5m_Q^2\langle g_s^3GGG\rangle}{3072\pi^4T^2} \int_{z_i}^{z_f}dz \int_{y_i}^{y_f}dy \, \frac{1}{zy}\delta\left(s-\widetilde{m}_{Q}^2\right)\nonumber\\
&&- \frac{m_Q^2\langle g_s^3GGG\rangle}{512\pi^4T^2} \int_{z_i}^{z_f}dz \int_{y_i}^{y_f}dy \, \frac{1-y-z}{zy^{2}}\delta\left(s-\widetilde{m}_{Q}^2\right)\, ,
\end{eqnarray}

\begin{eqnarray}
\rho^0_{QQQ,\frac{3}{2}}(s)&=&\frac{3}{32\pi^4}\int_{z_i}^{z_f}dz \int_{y_i}^{y_f}dy \, yz \, (s-\widetilde{m}_{Q}^2)(8s-3\widetilde{m}_{Q}^2) + \frac{9 m_Q^2}{32\pi^4}\int_{z_i}^{z_f}dz \int_{y_i}^{y_f}dy \, (s-\widetilde{m}_Q^2)\nonumber\\
&&-\frac{m_Q^2}{192\pi^2}\langle\frac{\alpha_{s}GG}{\pi}\rangle\int_{z_i}^{z_f}dz \int_{y_i}^{y_f}dy \, \frac{z(1+y-z)}{y^{3}} \left( 1+\frac{5s}{T^2}\right)\delta\left(s-\widetilde{m}_{Q}^2\right)\nonumber\\
&&-\frac{3m_Q^4}{64\pi^2T^2}\langle\frac{\alpha_{s}GG}{\pi}\rangle\int_{z_i}^{z_f}dz \int_{y_i}^{y_f}dy \, \frac{1}{y^{3}} \delta\left(s-\widetilde{m}_{Q}^2\right)\nonumber\\
&&+\frac{3}{32\pi^2}\langle\frac{\alpha_{s}GG}{\pi}\rangle\int_{z_i}^{z_f}dz \int_{y_i}^{y_f}dy \, \frac{z(1-y-z)}{y^{2}} \left[ 1+\frac{5s}{6}\delta\left(s-\widetilde{m}_{Q}^2\right)\right]\nonumber\\
&&+\frac{9m_Q^2}{64\pi^2}\langle\frac{\alpha_{s}GG}{\pi}\rangle\int_{z_i}^{z_f}dz \int_{y_i}^{y_f}dy \, \frac{1}{y^{2}}\delta\left(s-\widetilde{m}_{Q}^2\right)\nonumber\\
&&-\frac{3}{64\pi^2}\langle\frac{\alpha_{s}GG}{\pi}\rangle\int_{z_i}^{z_f}dz \int_{y_i}^{y_f}dy \, \left[ 1+\frac{7s}{18}\delta\left(s-\widetilde{m}_{Q}^2\right)\right]\nonumber\\
&&-\frac{1}{32\pi^2}\langle\frac{\alpha_{s}GG}{\pi}\rangle\int_{z_i}^{z_f}dz \int_{y_i}^{y_f}dy \, \frac{1-y-z}{z}\left[ 1+\frac{5s}{6}\delta\left(s-\widetilde{m}_{Q}^2\right)\right]\nonumber\\
&&-\frac{m_Q^2}{64\pi^2}\langle\frac{\alpha_{s}GG}{\pi}\rangle\int_{z_i}^{z_f}dz \int_{y_i}^{y_f}dy \, \frac{1}{zy}\delta\left(s-\widetilde{m}_{Q}^2\right)\nonumber\\
&&- \frac{m_Q^2\langle g_s^3GGG\rangle}{384\pi^4T^2}\int_{z_i}^{z_f}dz \int_{y_i}^{y_f}dy \, \frac{z(1+y-z)}{y^{4}}\left( 1-\frac{5s}{4T^2}\right)\delta\left(s-\widetilde{m}_{Q}^2\right)\nonumber\\
&&+ \frac{3m_Q^4\langle g_s^3GGG\rangle}{512\pi^4T^4} \int_{z_i}^{z_f}dz \int_{y_i}^{y_f}dy \, \frac{1}{y^{4}}\delta\left(s-\widetilde{m}_{Q}^2\right)\nonumber\\
&&- \frac{\langle g_s^3GGG\rangle}{1536\pi^4}\int_{z_i}^{z_f}dz \int_{y_i}^{y_f}dy \, \frac{z(3-2y-3z)}{y^{3}}\left( 1+\frac{5s}{T^2}\right)\delta\left(s-\widetilde{m}_{Q}^2\right)\nonumber\\
&&- \frac{9m_Q^2\langle g_s^3GGG\rangle}{512\pi^4T^2} \int_{z_i}^{z_f}dz \int_{y_i}^{y_f}dy \, \frac{1}{y^{3}}\delta\left(s-\widetilde{m}_{Q}^2\right)\nonumber\\
&&+\frac{\langle g_s^3GGG\rangle}{4608\pi^4} \int_{z_i}^{z_f}dz \int_{y_i}^{y_f}dy \, \frac{1-y-z}{y^{2}}\left(1+\frac{2ys}{T^2} \right)\left( 1+\frac{s}{T^2}\right)\delta\left(s-\widetilde{m}_{Q}^2\right)\nonumber\\
&&+ \frac{m_Q^2\langle g_s^3GGG\rangle}{1152\pi^4T^2}\int_{z_i}^{z_f}dz \int_{y_i}^{y_f}dy \, \frac{1}{y^{2}}\left( 1+\frac{7s}{4T^2}\right)\delta\left(s-\widetilde{m}_{Q}^2\right)\nonumber
\end{eqnarray}

\begin{eqnarray}
&&- \frac{m_Q^2\langle g_s^3GGG\rangle}{2304\pi^4T^2}\int_{z_i}^{z_f}dz \int_{y_i}^{y_f}dy \, \frac{1-y-z}{y^{3}}\left( 1-\frac{3s}{2T^2}\right)\delta\left(s-\widetilde{m}_{Q}^2\right)\nonumber\\
&&- \frac{m_Q^2\langle g_s^3GGG\rangle}{1152\pi^4T^2} \int_{z_i}^{z_f}dz \int_{y_i}^{y_f}dy \, \frac{1-y-z}{zy^{2}}\left( 1-\frac{5s}{4T^2}\right)\delta\left(s-\widetilde{m}_{Q}^2\right)\nonumber\\
&&+ \frac{m_Q^2\langle g_s^3GGG\rangle}{576\pi^4T^2} \int_{z_i}^{z_f}dz \int_{y_i}^{y_f}dy \, \frac{1-y}{zy^{2}}\delta\left(s-\widetilde{m}_{Q}^2\right)\nonumber\\
&&+\frac{7m_Q^4\langle g_s^3GGG\rangle }{4608\pi^4T^4} \int_{z_i}^{z_f}dz \int_{y_i}^{y_f}dy \, \frac{1}{zy^{3}}\delta\left(s-\widetilde{m}_{Q}^2\right)\nonumber\\
&&-\frac{11\langle g_s^3GGG\rangle}{3072\pi^4} \int_{z_i}^{z_f}dz \int_{y_i}^{y_f}dy \, \frac{1}{y}\left( 1+\frac{7s}{11T^2}\right)\delta\left(s-\widetilde{m}_{Q}^2\right)\nonumber\\
&&-\frac{\langle g_s^3GGG\rangle}{1024\pi^4}\int_{z_i}^{z_f}dz \int_{y_i}^{y_f}dy \, \frac{1-y-z}{y^{2}}\left( 1+\frac{11s}{3T^2}\right)\delta\left(s-\widetilde{m}_{Q}^2\right)\nonumber\\
&&- \frac{\langle g_s^3GGG\rangle}{3072\pi^4} \int_{z_i}^{z_f}dz \int_{y_i}^{y_f}dy \, \frac{1-y-z}{zy}\left( 1+\frac{5s}{T^2}\right)\delta\left(s-\widetilde{m}_{Q}^2\right)\nonumber\\
&&- \frac{5m_Q^2\langle g_s^3GGG\rangle}{768\pi^4T^2} \int_{z_i}^{z_f}dz \int_{y_i}^{y_f}dy \, \frac{1}{zy^{2}}\delta\left(s-\widetilde{m}_{Q}^2\right)\, ,
\end{eqnarray}

\begin{eqnarray}
\rho^1_{QQQ^\prime,\frac{3}{2}}(s)&=&\frac{3}{16\pi^4}\int_{z_i}^{z_f}dz \int_{y_i}^{y_f}dy \, yz(1-y-z) (s-\widetilde{m}_{Q}^2)(2s-\widetilde{m}_{Q}^2) \nonumber\\
&&+ \frac{3m_Q^2}{16\pi^4}\int_{z_i}^{z_f}dz \int_{y_i}^{y_f}dy \,z\, (s-\widetilde{m}_Q^2) \nonumber\\
&&-\frac{m_Q^2}{48\pi^2}\langle\frac{\alpha_{s}GG}{\pi}\rangle\int_{z_i}^{z_f}dz \int_{y_i}^{y_f}dy \, \frac{z(1-y-z)}{y^{2}} \left( 1+\frac{s}{T^2}\right)\delta\left(s-\widetilde{m}_{Q}^2\right) \nonumber\\
&&-\frac{m_Q^4}{48\pi^2T^2}\langle\frac{\alpha_{s}GG}{\pi}\rangle\int_{z_i}^{z_f}dz \int_{y_i}^{y_f}dy \, \frac{z}{y^{3}} \delta\left(s-\widetilde{m}_{Q}^2\right) \nonumber\\
&&+ \frac{m_Q^2}{16\pi^2}\langle\frac{\alpha_{s}GG}{\pi}\rangle\int_{z_i}^{z_f}dz \int_{y_i}^{y_f}dy \, \frac{z}{y^{2}}\delta\left(s-\widetilde{m}_{Q}^2\right) \nonumber\\
&&-\frac{m_{Q^\prime}^2}{96\pi^2}\langle\frac{\alpha_{s}GG}{\pi}\rangle\int_{z_i}^{z_f}dz \int_{y_i}^{y_f}dy \, \frac{y(1-y-z)}{z^{2}} \left( 1+\frac{s}{T^2}\right)\delta\left(s-\widetilde{m}_{Q}^2\right) \nonumber\\
&&-\frac{m_{Q^\prime}^2m_Q^2}{96\pi^2T^2}\langle\frac{\alpha_{s}GG}{\pi}\rangle\int_{z_i}^{z_f}dz \int_{y_i}^{y_f}dy \, \frac{1}{z^{2}} \delta\left(s-\widetilde{m}_{Q}^2\right) \nonumber\\
&&-\frac{1}{48\pi^2}\langle\frac{\alpha_{s}GG}{\pi}\rangle\int_{z_i}^{z_f}dz \int_{y_i}^{y_f}dy \, z \left[ 1+\frac{s}{4}\delta\left(s-\widetilde{m}_{Q}^2\right)\right] \, ,
\end{eqnarray}

\begin{eqnarray}
\rho^0_{QQQ^\prime,\frac{3}{2}}(s)&=&\frac{3}{32\pi^4}\int_{z_i}^{z_f}dz \int_{y_i}^{y_f}dy \, y(1-y-z) (s-\widetilde{m}_{Q}^2)(3s-\widetilde{m}_{Q}^2) \nonumber\\
&&+\frac{3m_Q^2}{16\pi^4}  \int_{z_i}^{z_f}dz \int_{y_i}^{y_f}dy \, (s-\widetilde{m}_Q^2) \nonumber\\
&&-\frac{m_Q^2}{48\pi^2T^2}\langle\frac{\alpha_{s}GG}{\pi}\rangle\int_{z_i}^{z_f}dz \int_{y_i}^{y_f}dy \, \frac{(1-y-z)}{y^{2}} \,s\, \delta\left(s-\widetilde{m}_{Q}^2\right) \nonumber
\end{eqnarray}

\begin{eqnarray}
&&- \frac{m_Q^4}{48\pi^2T^2} \langle\frac{\alpha_{s}GG}{\pi}\rangle\int_{z_i}^{z_f}dz \int_{y_i}^{y_f}dy \, \frac{1}{y^{3}} \delta\left(s-\widetilde{m}_{Q}^2\right)
\nonumber\\
&&+\frac{m_Q^2}{16\pi^2}\langle\frac{\alpha_{s}GG}{\pi}\rangle\int_{z_i}^{z_f}dz \int_{y_i}^{y_f}dy \, \frac{1}{y^{2}} \delta\left(s-\widetilde{m}_{Q}^2\right) \nonumber\\
&&-\frac{m_{Q^\prime}^2}{96\pi^2T^2}\langle\frac{\alpha_{s}GG}{\pi}\rangle\int_{z_i}^{z_f}dz \int_{y_i}^{y_f}dy \, \frac{y(1-y-z)}{z^{3}} \,s\, \delta\left(s-\widetilde{m}_{Q}^2\right) \nonumber\\
&&- \frac{m_{Q^\prime}^2m_Q^2}{96\pi^2T^2}\langle\frac{\alpha_{s}GG}{\pi}\rangle\int_{z_i}^{z_f}dz \int_{y_i}^{y_f}dy \, \frac{1}{z^{3}} \delta\left(s-\widetilde{m}_{Q}^2\right) \nonumber\\
&&+\frac{1}{32\pi^2}\langle\frac{\alpha_{s}GG}{\pi}\rangle\int_{z_i}^{z_f}dz \int_{y_i}^{y_f}dy \, \frac{y(1-y-z)}{z^{2}} \Big[ 1+s\delta\left(s-\widetilde{m}_{Q}^2\right)\Big] \nonumber\\
&&+\frac{m_Q^2}{32\pi^2}\langle\frac{\alpha_{s}GG}{\pi}\rangle\int_{z_i}^{z_f}dz \int_{y_i}^{y_f}dy \, \frac{1}{z^{2}} \delta\left(s-\widetilde{m}_{Q}^2\right) \nonumber\\
&&-\frac{1}{64\pi^2}\langle\frac{\alpha_{s}GG}{\pi}\rangle\int_{z_i}^{z_f}dz \int_{y_i}^{y_f}dy \, \left[ 1+\frac{s}{3}\delta\left(s-\widetilde{m}_{Q}^2\right)\right] \, ,
\end{eqnarray}

\begin{eqnarray}
\rho^1_{QQQ^\prime,\frac{1}{2}}(s)&=&\frac{3}{8\pi^4}\int_{z_i}^{z_f}dz \int_{y_i}^{y_f}dy \, yz(1-y-z) (s-\widetilde{m}_{Q}^2)(5s-3\widetilde{m}_{Q}^2) \nonumber\\
&&+\frac{3m_Q^2}{8\pi^4} \int_{z_i}^{z_f}dz \int_{y_i}^{y_f}dy \,z\, (s-\widetilde{m}_Q^2) \nonumber\\
&&-\frac{m_Q^2}{6\pi^2}\langle\frac{\alpha_{s}GG}{\pi}\rangle\int_{z_i}^{z_f}dz \int_{y_i}^{y_f}dy \, \frac{z(1-y-z)}{y^{2}} \left( 1+\frac{s}{2T^2}\right)\delta\left(s-\widetilde{m}_{Q}^2\right) \nonumber\\
&&-\frac{m_Q^4}{24\pi^2T^2}\langle\frac{\alpha_{s}GG}{\pi}\rangle\int_{z_i}^{z_f}dz \int_{y_i}^{y_f}dy \, \frac{z}{y^{3}} \delta\left(s-\widetilde{m}_{Q}^2\right) \nonumber\\
&&+\frac{m_Q^2}{8\pi^2}\langle\frac{\alpha_{s}GG}{\pi}\rangle\int_{z_i}^{z_f}dz \int_{y_i}^{y_f}dy \, \frac{z}{y^{2}} \delta\left(s-\widetilde{m}_{Q}^2\right) \nonumber\\
&&-\frac{m_{Q^\prime}^2}{12\pi^2}\langle\frac{\alpha_{s}GG}{\pi}\rangle\int_{z_i}^{z_f}dz \int_{y_i}^{y_f}dy \, \frac{y(1-y-z)}{z^{2}} \left( 1+\frac{s}{2T^2}\right)\delta\left(s-\widetilde{m}_{Q}^2\right) \nonumber\\
&&-\frac{m_{Q^\prime}^2m_Q^2}{48\pi^2T^2}\langle\frac{\alpha_{s}GG}{\pi}\rangle\int_{z_i}^{z_f}dz \int_{y_i}^{y_f}dy \, \frac{1}{z^{2}} \delta\left(s-\widetilde{m}_{Q}^2\right) \nonumber\\
&&+\frac{3}{16\pi^2}\langle\frac{\alpha_{s}GG}{\pi}\rangle\int_{z_i}^{z_f}dz \int_{y_i}^{y_f}dy \, (1-y-z) \left[ 1+\frac{s}{3}\delta\left(s-\widetilde{m}_{Q}^2\right)\right] \nonumber\\
&&+\frac{m_Q^2}{16\pi^2}\langle\frac{\alpha_{s}GG}{\pi}\rangle\int_{z_i}^{z_f}dz \int_{y_i}^{y_f}dy \, \frac{1}{y} \delta\left(s-\widetilde{m}_{Q}^2\right) \, ,
\end{eqnarray}

\begin{eqnarray}
\rho^0_{QQQ^\prime,\frac{1}{2}}(s)&=&\frac{3}{8\pi^4}\int_{z_i}^{z_f}dz \int_{y_i}^{y_f}dy \, y(1-y-z) (s-\widetilde{m}_{Q}^2)(2s-\widetilde{m}_{Q}^2) \nonumber\\
&&+\frac{3m_Q^2}{4\pi^4}\int_{z_i}^{z_f}dz \int_{y_i}^{y_f}dy \, (s-\widetilde{m}_Q^2) \nonumber\\
&&-\frac{m_Q^2}{24\pi^2}\langle\frac{\alpha_{s}GG}{\pi}\rangle\int_{z_i}^{z_f}dz \int_{y_i}^{y_f}dy \, \frac{(1-y-z)}{y^{2}} \left( 1+\frac{s}{T^2}\right)\delta\left(s-\widetilde{m}_{Q}^2\right) \nonumber
\end{eqnarray}

\begin{eqnarray}
&&-\frac{m_Q^4}{12\pi^2T^2}\langle\frac{\alpha_{s}GG}{\pi}\rangle\int_{z_i}^{z_f}dz \int_{y_i}^{y_f}dy \, \frac{1}{y^{3}} \delta\left(s-\widetilde{m}_{Q}^2\right) \nonumber\\
&&+ \frac{m_Q^2}{4\pi^2}\langle\frac{\alpha_{s}GG}{\pi}\rangle\int_{z_i}^{z_f}dz \int_{y_i}^{y_f}dy \, \frac{1}{y^{2}}\delta\left(s-\widetilde{m}_{Q}^2\right) \nonumber\\
&&-\frac{m_{Q^\prime}^2}{48\pi^2}\langle\frac{\alpha_{s}GG}{\pi}\rangle\int_{z_i}^{z_f}dz \int_{y_i}^{y_f}dy \, \frac{y(1-y-z)}{z^{3}} \left( 1+\frac{s}{T^2}\right)\delta\left(s-\widetilde{m}_{Q}^2\right) \nonumber\\
&&-\frac{m_{Q^\prime}^2m_Q^2}{24\pi^2T^2} \langle\frac{\alpha_{s}GG}{\pi}\rangle\int_{z_i}^{z_f}dz \int_{y_i}^{y_f}dy \, \frac{1}{z^{3}} \delta\left(s-\widetilde{m}_{Q}^2\right) \nonumber\\
&&+\frac{1}{8\pi^2}\langle\frac{\alpha_{s}GG}{\pi}\rangle\int_{z_i}^{z_f}dz \int_{y_i}^{y_f}dy \, \frac{y(1-y-z)}{z^{2}} \left[ 1+\frac{s}{2}\delta\left(s-\widetilde{m}_{Q}^2\right)\right] \nonumber\\
&&+\frac{m_Q^2}{8\pi^2}\langle\frac{\alpha_{s}GG}{\pi}\rangle\int_{z_i}^{z_f}dz \int_{y_i}^{y_f}dy \, \frac{1}{z^{2}} \delta\left(s-\widetilde{m}_{Q}^2\right) \nonumber\\
&&-\frac{1}{16\pi^2}\langle\frac{\alpha_{s}GG}{\pi}\rangle\int_{z_i}^{z_f}dz \int_{y_i}^{y_f}dy \, \left[ 1+\frac{s}{2}\delta\left(s-\widetilde{m}_{Q}^2\right)\right] \nonumber\\
&&+\frac{1}{8\pi^2}\langle\frac{\alpha_{s}GG}{\pi}\rangle\int_{z_i}^{z_f}dz \int_{y_i}^{y_f}dy \,\frac{(1-y-z)}{z} \left[ 1+\frac{s}{2}\delta\left(s-\widetilde{m}_{Q}^2\right)\right] \nonumber\\
&&+\frac{m_Q^2}{8\pi^2}\langle\frac{\alpha_{s}GG}{\pi}\rangle\int_{z_i}^{z_f}dz \int_{y_i}^{y_f}dy \, \frac{1}{zy} \delta\left(s-\widetilde{m}_{Q}^2\right) \, ,
\end{eqnarray}
$\widetilde{m}_Q^2=\frac{m_Q^2}{y}+\frac{m_Q^2}{1-y-z}+\frac{m_{Q^\prime}^2}{z}$,
\begin{eqnarray}
z_{i/f}&=&\frac{s+m_{Q^\prime}^2-4m_Q^2\mp\sqrt{(s+m_{Q^\prime}^2-4m_Q^2)^2-4sm_{Q^\prime}^2}}{2s}\, ,\nonumber\\
y_{i/f}&=&\frac{1-z \mp \sqrt{(1-z)^2-4z(1-z)m_Q^2/(zs-m_{Q^\prime}^2)}}{2}\, ,
\end{eqnarray}
in the case of the $QQQ$ baryon states, we set  $m_{Q^\prime}^2=m_{Q}^2$. When the $\delta\left(s-\widetilde{m}_{Q}^2\right)$ functions appear, $\int_{z_i}^{z_f}dz \int_{y_i}^{y_f}dy \to \int_{0}^{1}dz \int^{1-z}_{0}dy  $.

We  differentiate  Eq.\eqref{QCDSR} with respect to  $\tau=\frac{1}{T^2}$, then eliminate the
 pole residues $\lambda_{+}$ and obtain the QCD sum rules for  the masses of the triply-heavy baryon states,
 \begin{eqnarray}\label{QCDSR-mass}
 M^2_{+} &=& \frac{-\frac{d}{d\tau}\int_{\Delta^2}^{s_0}ds \,\left[\sqrt{s}\rho^1_{QCD}(s)+ \rho^0_{QCD}(s)\right]\exp\left( -s\tau\right)}{\int_{\Delta^2}^{s_0}ds\, \left[\sqrt{s}\rho^1_{QCD}(s)+ \rho^0_{QCD}(s)\right]\exp\left( -s\tau\right)}\,  .
\end{eqnarray}

\section{Numerical results and discussions}
We take  the standard values of the  gluon condensates
 $\langle \frac{\alpha_sGG}{\pi}\rangle=0.012\pm0.004\,\rm{GeV}^4$ and the three-gluon condensates $\langle g_s^3GGG\rangle=0.045\pm0.014\,\rm{GeV}^6$
\cite{SVZ79,Reinders85,ColangeloReview}, and  take the $\overline{MS}$ masses of the heavy quarks $m_{c}(m_c)=(1.275\pm0.025)\,\rm{GeV}$
 and $m_{b}(m_b)=(4.18\pm0.03)\,\rm{GeV}$
 from the Particle Data Group \cite{PDG}, which work well in studying the doubly-heavy baryon states \cite{Wang-cc-baryon-penta}, hidden-charm tetraquark states \cite{WangZG-Hiddencharm, WangZG-Vector-tetra}, hidden-bottom tetraquark states \cite{ WangZG-Hiddenbottom}, hidden-charm pentaquark states \cite{WangZG-hiddencharm-penta}, fully-charmed tetraquark states \cite{WangZG-QQQQ}, and perform a new analysis.
Furthermore,  we take into account
the energy-scale dependence of  the  $\overline{MS}$ masses according to  the renormalization group equation,
 \begin{eqnarray}
 m_c(\mu)&=&m_c(m_c)\left[\frac{\alpha_{s}(\mu)}{\alpha_{s}(m_c)}\right]^{\frac{12}{33-2n_f}} \, ,\nonumber\\
m_b(\mu)&=&m_b(m_b)\left[\frac{\alpha_{s}(\mu)}{\alpha_{s}(m_b)}\right]^{\frac{12}{33-2n_f}} \, ,\nonumber\\
\alpha_s(\mu)&=&\frac{1}{b_0t}\left[1-\frac{b_1}{b_0^2}\frac{\log t}{t} +\frac{b_1^2(\log^2{t}-\log{t}-1)+b_0b_2}{b_0^4t^2}\right]\, ,
\end{eqnarray}
  where $t=\log \frac{\mu^2}{\Lambda^2}$, $b_0=\frac{33-2n_f}{12\pi}$, $b_1=\frac{153-19n_f}{24\pi^2}$, $b_2=\frac{2857-\frac{5033}{9}n_f+\frac{325}{27}n_f^2}{128\pi^3}$,  $\Lambda=210\,\rm{MeV}$, $292\,\rm{MeV}$  and  $332\,\rm{MeV}$ for the flavors  $n_f=5$, $4$ and $3$, respectively  \cite{PDG}.
For the $ccb$, $bbc$ and $bbb$ baryon states, we choose the flavor numbers $n_f=5$, for the $ccc$ baryon state, we choose the flavor numbers $n_f=4$, and choose the optimal  energy scales $\mu$ to obtain stable QCD sum rules in different channels and to enhance the pole contributions in a consistent way, while in previous works, the heavy quark masses were just taken as mass parameters, had constant values in all the QCD sum rules  \cite{QCDSR-1,QCDSR-2,QCDSR-3,QCDSR-4}.

As long as the continuum threshold parameters $s_0$ are concerned, we should choose suitable values  to avoid contaminations from the first radial excited states of the triply-heavy baryon states,   and can borrow some ideas from the experimental data on the  conventional charmonium (bottomonium) states and the charmonium-like states.
The   energy gaps between the ground states and the first radial excited states are $M_{\psi^\prime}-M_{J/\psi}=589\,\rm{MeV}$ and $M_{\Upsilon^\prime}-M_{\Upsilon}=563\,\rm{MeV}$ from the Particle Data Group \cite{PDG},
$M_{B_c^{*\prime}}-M_{B^*_c}=567 \,\rm{MeV}$ from the CMS collaboration \cite{CMS-Bc-1902},  $M_{B_c^{*\prime}}-M_{B^*_c}=566 \,\rm{MeV}$ from the LHCb collaboration \cite{LHCb-Bc-1904},  $M_{Z_c(4430)}-M_{Z_c(3900)}=591\,\rm{MeV}$, $M_{X(4500)}-M_{X(3915)}=588\,\rm{MeV}$  from the Particle Data Group \cite{PDG},
 $M_{Z_c(4600)}-M_{Z_c(4020)}=576\,\rm{MeV}$ from the LHCb collaboration \cite{LHCb-Z4600}.
We usually assign the $Z_c^\pm(3900)$ and $Z_c^\pm(4430)$  to be the ground state and the first radial excited state of the axialvector tetraquark states respectively \cite{Maiani-Z4430-1405},
assign the $X(3915)$ and  $X(4500)$  to be   the ground state and the first radial excited state of the  scalar  tetraquark states  respectively \cite{Lebed-X3915,WangZG-X4500}, assign  the $Z_c^\pm(4020)$ and $Z_c^\pm(4600)$ to be the ground state  and the first radial excited state of the axialvector tetraquark states respectively with different quark structures from that of the $Z_c^\pm(3900)$ and $Z_c^\pm(4430)$ \cite{WangZG-Hiddencharm,ChenHX-Z4600-A}.

In the present work,  we can take  the experimental data from the Particle Data Group, CMS and LHCb collaborations as input parameters and choose the continuum threshold parameters as   $\sqrt{s_0}=M_{\Omega/\Omega^*}+(0.50\sim0.55)\,\rm{GeV}$ as a constraint to study the triply-heavy  baryon  states with the QCD sum rules, furthermore, we add an uncertainty $\delta\sqrt{s_0}=\pm0.1\,\rm{GeV}$ as we usually do in estimating  the uncertainties from the continuum threshold parameters in the QCD sum rules.
We vary the energy scales of the QCD spectral densities, the continuum threshold parameters and the Borel parameters to satisfy
the  two basic    criteria of the QCD sum rules, i.e. the ground state  dominance  at the hadron  side and the operator product expansion  convergence  at the QCD side.
 After trial and error, we obtain the ideal energy scales of the QCD spectral densities, the Borel parameters  $T^2$ and the continuum threshold parameters $s_0$, therefore  the pole contributions of the ground state triply-heavy baryon  states, see Table \ref{Borel-mass}.

\begin{table}
\begin{center}
\begin{tabular}{|c|c|c|c|c|c|c|c|}\hline\hline
                        &$T^2(\rm{GeV}^2)$   &$\sqrt{s_0}(\rm{GeV})$ &$\mu(\rm{GeV})$    &pole          &$M(\rm{GeV})$   &$\lambda(10^{-1}\rm{GeV}^3)$ \\ \hline

$ccc({\frac{3}{2}}^+)$  &$3.1-4.1$           &$5.35\pm0.10$          &$1.2$              &$(56-83)\%$   &$4.81\pm0.10$   &$(2.08\pm0.31)$   \\ \hline

$ccb({\frac{3}{2}}^+)$  &$4.7-5.7$           &$8.55\pm0.10$          &$2.1$              &$(65-85)\%$   &$8.03\pm0.08$   &$(2.25\pm0.25)$   \\ \hline

$ccb({\frac{1}{2}}^+)$  &$4.9-5.9$           &$8.55\pm0.10$          &$2.0$              &$(64-84)\%$   &$8.02\pm0.08$   &$(4.30\pm0.47)$   \\ \hline

$bbc({\frac{3}{2}}^+)$  &$6.4-7.4$           &$11.75\pm0.10$         &$2.2$              &$(65-83)\%$   &$11.23\pm0.08$  &$(3.24\pm0.46)$   \\ \hline

$bbc({\frac{1}{2}}^+)$  &$6.3-7.3$           &$11.75\pm0.10$         &$2.2$              &$(65-84)\%$   &$11.22\pm0.08$  &$(5.65\pm0.81)$   \\ \hline

$bbb({\frac{3}{2}}^+)$  &$8.6-9.6$           &$14.95\pm0.10$         &$2.5$              &$(66-83)\%$   &$14.43\pm0.09$  &$(9.42\pm1.39)$   \\ \hline\hline
\end{tabular}
\end{center}
\caption{ The Borel windows, continuum threshold parameters, energy scales of the QCD spectral densities,  pole contributions,   masses and pole residues for the
triply-heavy baryon  states. } \label{Borel-mass}
\end{table}

\begin{figure}
 \centering
 \includegraphics[totalheight=5cm,width=7cm]{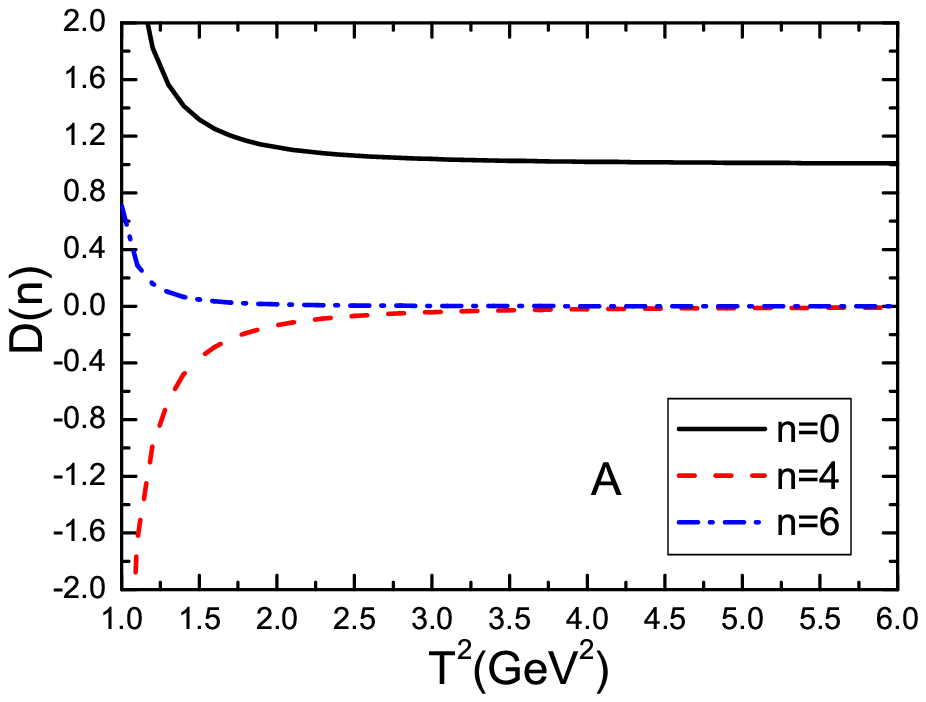}
  \includegraphics[totalheight=5cm,width=7cm]{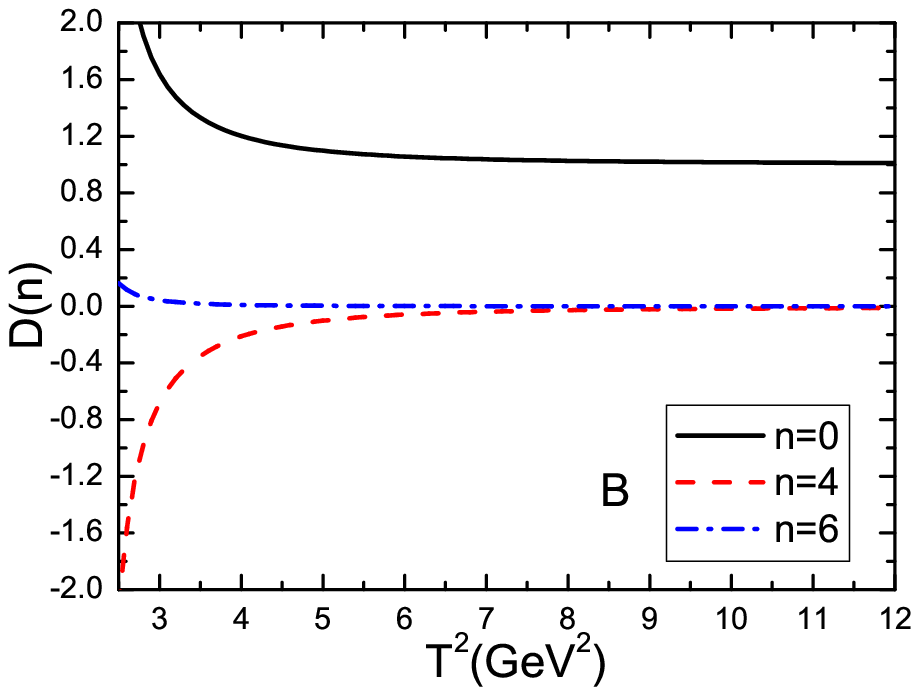}
 \caption{ The contributions of the vacuum condensates  with variations  of the Borel parameters  $T^2$, where the $A$ and $B$ denote the baryon states $ccc({\frac{3}{2}}^+)$ and $bbb({\frac{3}{2}}^+)$, respectively, the $n=0$, $4$, $6$ denotes the dimensions of the vacuum condensates.   }\label{fr-fig}
\end{figure}

 In the Borel windows, the pole contributions are about $(60-80)\%$,  the pole dominance is well satisfied. In Fig.\ref{fr-fig}, we plot the contributions of the perturbative terms, the gluon condensates and the  three-gluon condensates with variations of the Borel parameters for the central values of the continuum threshold parameters shown in Table \ref{Borel-mass} in the QCD sum rules for the $ccc$ and $bbb$ baryon states. From the figure, we can see that the main contributions come from the perturbative terms, the gluon condensates  play a minor important role, and the three-gluon condensates play a tiny role in the Borel windows. The Borel parameters have the relation $T_{bbb}^2> T^2_{bbc}>T^2_{ccb}>T^2_{ccc}$, we add the subscripts $bbb$, $bbc$, $ccb$ and $ccc$ to denote the corresponding QCD sum rules. From Fig.\ref{fr-fig}, we can see that the contributions from the three-gluon condensates decrease quickly with the increase of the Borel parameters, at the region $T^2\geq 1.5\, \rm{GeV}^2$ in the $ccc$ channel and at the region $T^2\geq 3.0\, \rm{GeV}^2$ in the $bbb$ channel, the
  contributions from the three-gluon condensates reach zero and can be neglected safely.
  On the other hand, the Borel window  $T^2_{ccc}>3.0\,\rm{GeV}^2$, so we can neglect the three-gluon condensates in the QCD sum rules for the $ccb$ and $bbc$ baryon states without impairing  the predictive ability.  The operator product expansion is well convergent. Although the three-gluon condensates play a tiny role in the Borel windows and can be neglected in the Borel windows, we take them into account to obtain the values $T^2\geq 1.5\, \rm{GeV}^2$ or $T^2\geq 3.0\, \rm{GeV}^2$, the calculations are non-trivial.

\begin{figure}
 \centering
 \includegraphics[totalheight=5cm,width=7cm]{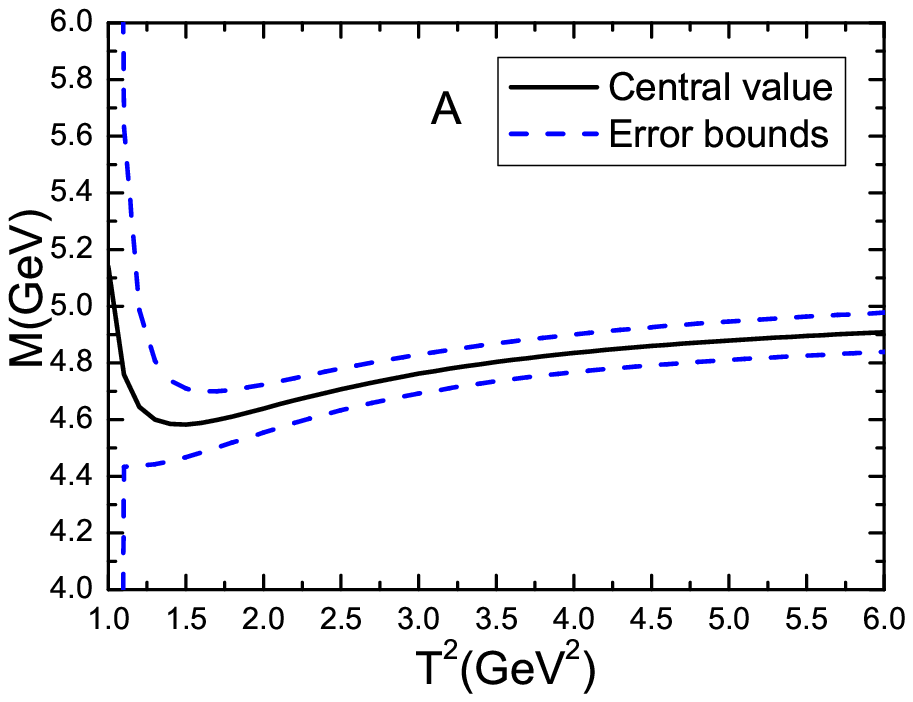}
 \includegraphics[totalheight=5cm,width=7cm]{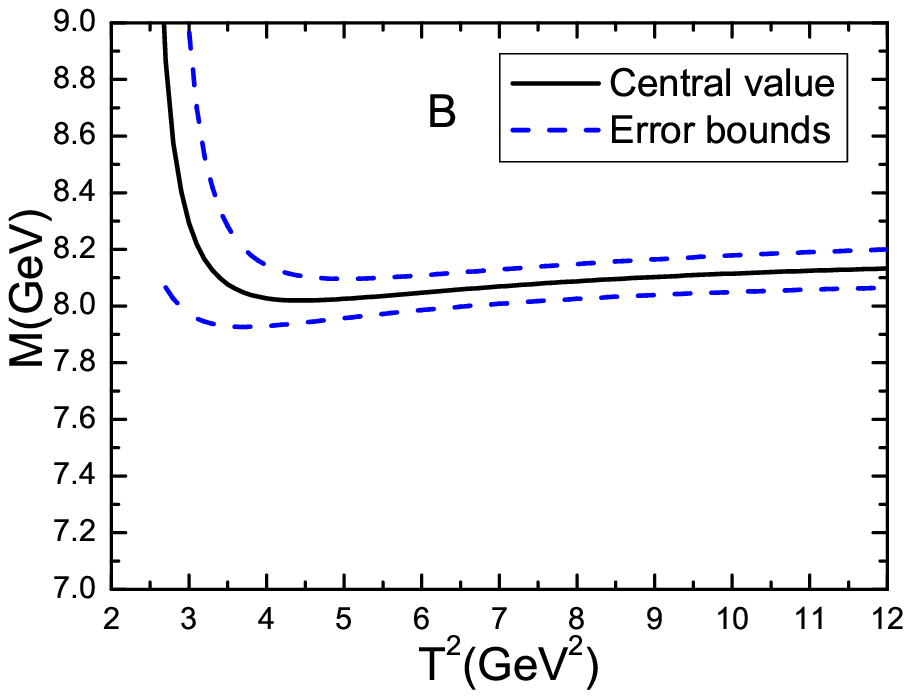}
 \includegraphics[totalheight=5cm,width=7cm]{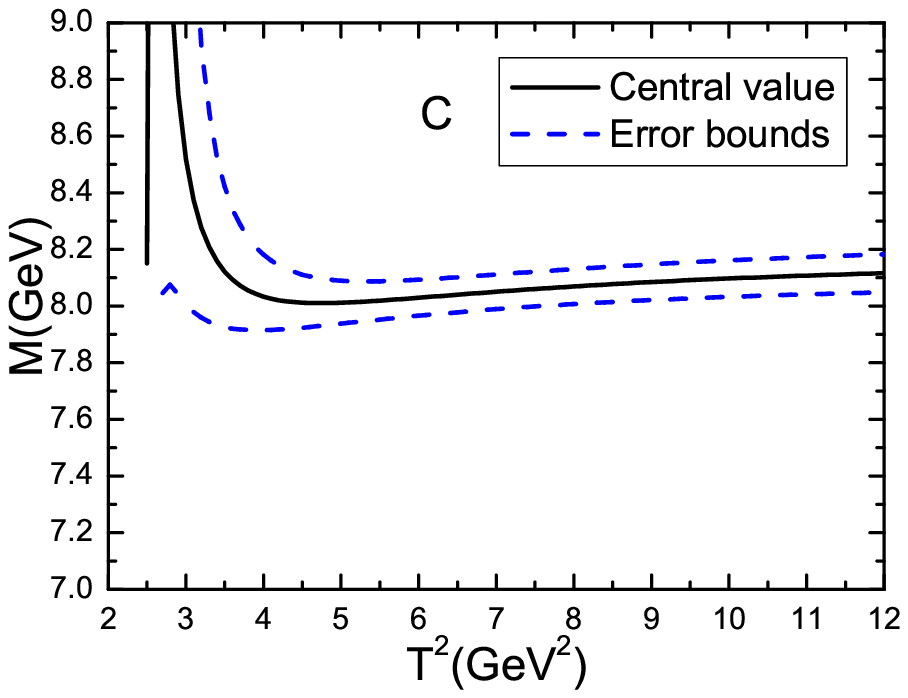}
 \includegraphics[totalheight=5cm,width=7cm]{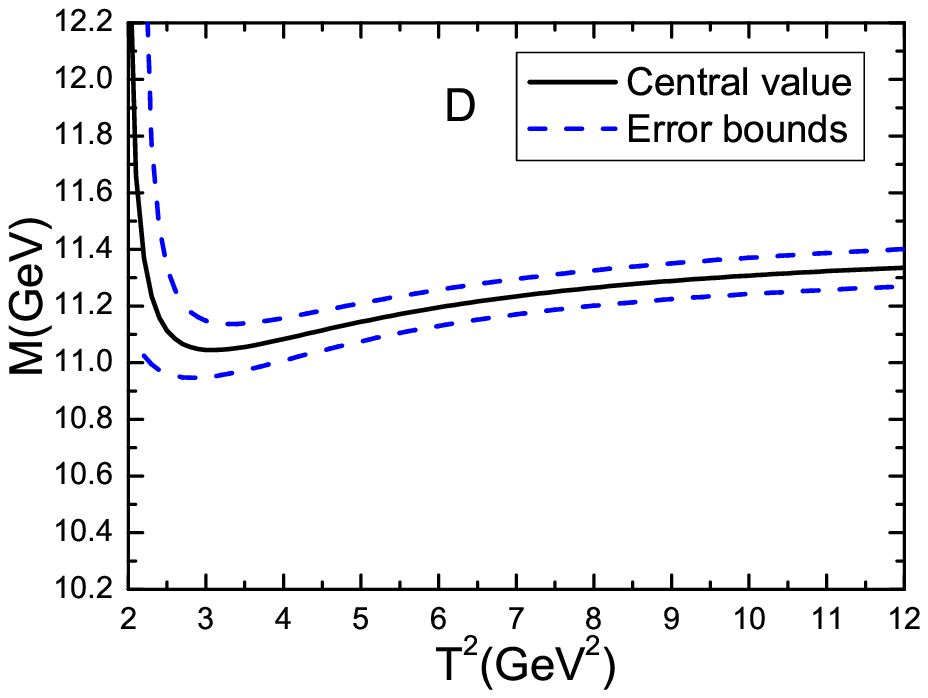}
 \includegraphics[totalheight=5cm,width=7cm]{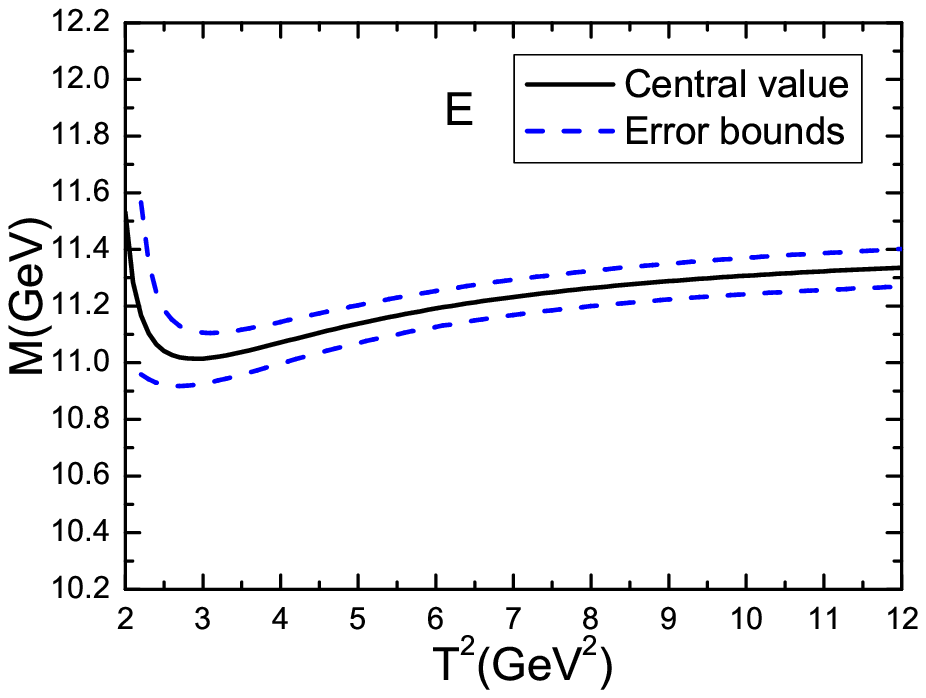}
 \includegraphics[totalheight=5cm,width=7cm]{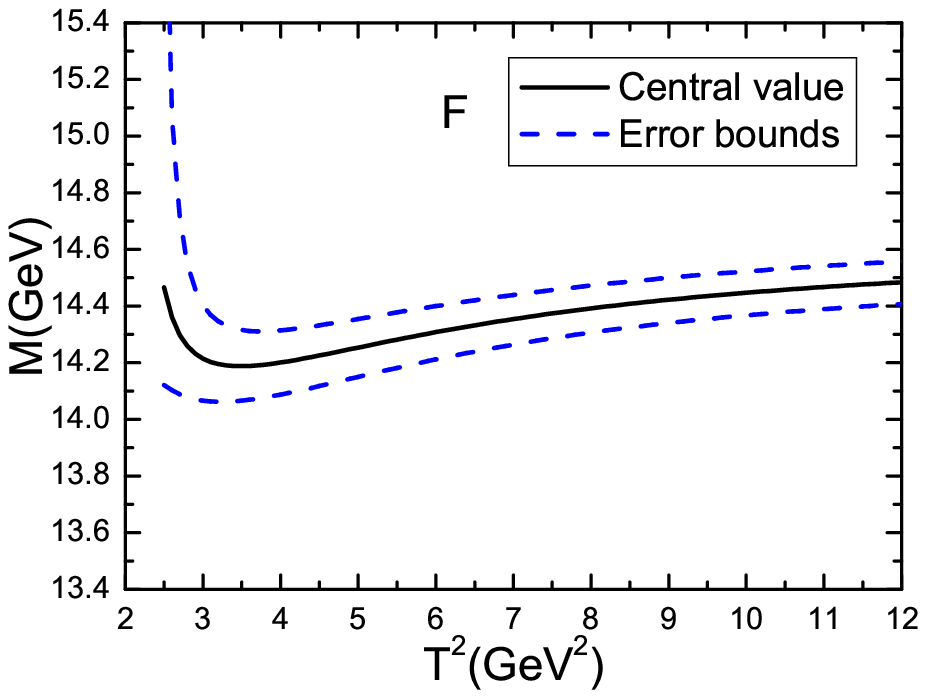}
 \caption{ The masses of the triply-heavy baryon  states with variations  of the Borel parameters  $T^2$, where the $A$, $B$, $C$, $D$, $E$ and $F$ denote the baryon states $ccc({\frac{3}{2}}^+)$, $ccb({\frac{3}{2}}^+)$, $ccb({\frac{1}{2}}^+)$, $bbc({\frac{3}{2}}^+)$, $bbc({\frac{1}{2}}^+)$ and
$bbb({\frac{3}{2}}^+)$, respectively.   }\label{mass-fig}
\end{figure}

Now we take  into account all uncertainties of the input parameters, and obtain the values of the masses and pole residues of
 the triply-heavy baryon states, which are  shown explicitly in Table \ref{Borel-mass} and Fig.\ref{mass-fig}.
 In Fig.\ref{mass-fig}, we plot the masses  of the  triply-heavy baryon states with variations  of the Borel parameters $T^2$ in  much large  ranges than the Borel windows.  From the figure, we can see that  there appear  platforms in the Borel windows indeed, the uncertainties originate from the Borel parameters are very small,    it is reliable to extract the triply-heavy baryon masses.

\begin{table}
\begin{center}
\begin{tabular}{|c|c|c|c|c|c|c|c|}\hline\hline
                    &$ccc({\frac{3}{2}}^+)$ &$ccb({\frac{3}{2}}^+)$ &$ccb({\frac{1}{2}}^+)$ &$bbc({\frac{3}{2}}^+)$ &$bbc({\frac{1}{2}}^+)$ &$bbb({\frac{3}{2}}^+)$ \\ \hline

This Work           &$4.81\pm0.10$          &$8.03\pm0.08$          &$8.02\pm0.08$          &$11.23\pm0.08$         &$11.22\pm0.08$         &$14.43\pm0.09$  \\ \hline

\cite{LQCD-1}       &$4.763$                &                       &                       &                       &                       &              \\ \hline

\cite{LQCD-2}       &$4.796$                &$8.037$                &$8.007$                &$11.229$               &$11.195$               &$14.366$     \\ \hline

\cite{LQCD-3}       &$4.769$                &                       &                       &                       &                       &              \\ \hline

\cite{LQCD-4}       &$4.789$                &                       &                       &                       &                       &              \\ \hline

\cite{LQCD-5}       &$4.761$                &                       &                       &                       &                       &              \\ \hline

\cite{LQCD-6}       &$4.734$                &                       &                       &                       &                       &              \\ \hline

\cite{LQCD-7}       &                       &                       &                       &                       &                       &$14.371$     \\ \hline

\cite{LQCD-8}       &                       &$8.026$                &$8.005$                &$11.211$               &$11.194$               &   \\ \hline

\cite{QCDSR-1}      &$4.67\pm 0.15$         &$7.45\pm 0.16$         &$7.41\pm 0.13$         &$10.54\pm 0.11$        &$10.30\pm 0.10$        &$13.28\pm 0.10$   \\ \hline

\cite{QCDSR-2}      &$4.72\pm0.12$          &$8.07\pm0.10$          &                       &$11.35\pm0.15$         &                       &$14.30\pm0.20$   \\ \hline

\cite{QCDSR-3}      &                       &                       &$8.50\pm 0.12$         &                       &$11.73\pm 0.16$        &   \\ \hline

\cite{QCDSR-4}      &$4.99 \pm 0.14$        &$8.23\pm 0.13$         &$8.23 \pm 0.13$        &$11.49 \pm 0.11$       &$11.50\pm 0.11$        &$14.83\pm 0.10$   \\ \hline

\cite{PQM-1}        &$4.965$                &$8.265$                &$8.245$                &$11.554$               &$11.535$               &$14.834$     \\ \hline

\cite{PQM-2}        &$4.798$                &$8.023$                &$8.004$                &$11.221$               &$11.200$               &$14.396$     \\ \hline

\cite{PQM-3}        &$4.763$                &                       &                       &                       &                       &$14.371$        \\ \hline

\cite{PQM-4}        &$4.760$                &$8.032$                &$7.999$                &$11.287$               &$11.274$               &$14.370$        \\ \hline

\cite{PQM-5}        &$4.799$                &$8.019$                &                       &$11.217$               &                       &$14.398$   \\ \hline

\cite{PQM-6}        &$4.76$                 &$7.98$                 &$7.98$                 &$11.19$                &$11.19$                &$14.37$        \\ \hline

\cite{PQM-7}        &$4.777$                &$8.005$                &$7.984$                &$11.163$               &$11.139$               &$14.276$        \\ \hline

\cite{PQM-8}        &$4.79$                 &$8.03$                 &                       &$11.20$                &                       &$14.30$        \\ \hline

\cite{PQM-9}        &$4.803$                &$8.025$                &$8.018$                &$11.287$               &$11.280$               &$14.569$        \\ \hline

\cite{PQM-10}       &$4.806$                &                       &                       &                       &                       &$14.496$   \\ \hline

\cite{PQM-11}       &$4.897$                &$8.273$                &$8.262$                &$11.589$               &$11.546$               &$14.688$        \\ \hline

\cite{PQM-12}       &$4.773$                &                       &                       &                       &                       &    \\ \hline

\cite{PQM-13}       &$4.828$                &                       &                       &                       &                       &$14.432$   \\ \hline

\cite{PQM-14}       &$4.900$                &$8.140$                &                       &$10.890$               &                       &$14.500$        \\ \hline

\cite{PQM-15}       &$4.799$                &$8.046$                &$8.018$                &$11.245$               &$11.214$               &$14.398$        \\ \hline

\cite{PQM-16}       &$4.798$                &                       &$8.018$                &                       &$11.215$               &$14.398$        \\ \hline

\cite{Faddeev-1}    &$4.760$                &$7.963$                &$7.867$                &$11.167$               &$11.077$               &$14.370$   \\ \hline

\cite{Faddeev-2}    &$4.799$                &                       &                       &                       &                       &$14.244$   \\ \hline

\cite{Faddeev-3}    &$5.00$                 &                       &$8.19$                 &                       &                       &$14.57$   \\ \hline

\cite{Faddeev-4}    &$4.93$                 &$8.03$                 &$8.01$                 &$11.12$                &$11.09$                &$14.23$   \\ \hline

\cite{Regge-1}      &$4.834$                &                       &                       &                       &                       &     \\ \hline

\cite{Regge-2}      &                       &                       &                       &                       &                       &$14.788$      \\ \hline

 \hline\hline
\end{tabular}
\end{center}
\caption{ The masses of the triply-heavy baryon states from different theoretical approaches, where the unit is GeV. } \label{QQQ-mass}
\end{table}

In Table \ref{QQQ-mass}, we also present the predictions of the triply-heavy baryon masses from the lattice QCD \cite{LQCD-1,LQCD-2,LQCD-3,LQCD-4,LQCD-5,LQCD-6,LQCD-7,LQCD-8}, the QCD sum rules \cite{QCDSR-1,QCDSR-2,QCDSR-3,QCDSR-4}, various potential quark models \cite{PQM-1,PQM-2,PQM-3,PQM-4,PQM-5,PQM-6,PQM-7,PQM-8,PQM-9,PQM-10,PQM-11,PQM-12,PQM-13,PQM-14,PQM-15,PQM-16}, the Fadeev equation \cite{Faddeev-1,Faddeev-2,Faddeev-3,Faddeev-4}, the Regge trajectories \cite{Regge-1,Regge-2}. From the Table, we can see that the predicted masses are $M_{ccc,{\frac{3}{2}}^+}=(4.7\sim5.0)\,\rm{GeV}$, $M_{ccb,{\frac{3}{2}}^+}=(7.5\sim8.3)\,\rm{GeV}$, $M_{ccb,{\frac{1}{2}}^+}=(7.4\sim8.5)\,\rm{GeV}$,
$M_{bbc,{\frac{3}{2}}^+}=(10.5\sim11.6)\,\rm{GeV}$, $M_{bbc,{\frac{1}{2}}^+}=(10.3\sim11.7)\,\rm{GeV}$, $M_{bbb,{\frac{3}{2}}^+}=(13.3\sim14.8)\,\rm{GeV}$ from the previous works, the present predictions $M_{ccc,{\frac{3}{2}}^+}=(4.81\pm0.10)\,\rm{GeV}$, $M_{ccb,{\frac{3}{2}}^+}=(8.03\pm0.08)\,\rm{GeV}$, $M_{ccb,{\frac{1}{2}}^+}=(8.02\pm0.08)\,\rm{GeV}$,
$M_{bbc,{\frac{3}{2}}^+}=(11.23\pm0.08)\,\rm{GeV}$, $M_{bbc,{\frac{1}{2}}^+}=(11.22\pm0.08)\,\rm{GeV}$, $M_{bbb,{\frac{3}{2}}^+}=(14.43\pm0.09)\,\rm{GeV}$, which are compatible with them, but with refined and more robust values compared to the previous  calculations based on the QCD sum rules \cite{QCDSR-1,QCDSR-2,QCDSR-3,QCDSR-4}.

\section{Conclusion}
In this article, we restudy the ground state triply-heavy baryon states with the QCD sum rules by carrying  out the operator product expansion up to the vacuum condensates of dimension 6 in a consistent way and performing  a novel analysis. It is the first time to take into account the three-gluon condensates in the QCD sum rules for the triply-heavy baryon states. In calculations,  we choose the $\overline{MS}$ masses of the heavy quarks, which work well in studying the doubly-heavy baryon states, hidden-charm tetraquark states, hidden-bottom tetraquark states, hidden-charm pentaquark states, fully-charmed tetraquark states, and vary the energy  scales to select the optimal values so as to obtain more stable QCD sum rules and enhance the pole contributions. The present predictions of the triply-heavy baryon masses are compatible with the existing theoretical calculations but with refined and more robust values compared to the previous  calculations based on the QCD sum rules, which can be confronted to the experimental data in the future to make contributions to the mass spectrum of the heavy baryon states.

\section*{Acknowledgements}
This  work is supported by National Natural Science Foundation, Grant Number  11775079.

\end{document}